\documentstyle[12pt,moriond,epsfig]{article}
\begin{document}
\newcommand{\etal}{{\em et al.}}
\heading{THE STAR FORMATION RATE OF THE LOCAL \\
	UNIVERSE FROM THE KPNO \\
	INTERNATIONAL SPECTROSCOPIC SURVEY}

\author{Caryl Gronwall}{Wesleyan University, Middletown, CT USA}

\begin{moriondabstract}

The KPNO International Spectroscopic Survey (KISS) is a wide-field
survey for extragalactic emission-line objects being carried out with
the Burrell Schmidt at Kitt Peak. The critical difference between
this survey and classical objective-prism surveys for active galaxies
is the use of a CCD detector which allows us to probe to much fainter
magnitudes than previous photographic surveys.  We have discovered
1126 emission-line galaxy (ELG) candidates in our H$\alpha$ selected
survey area of $\sim 68$ square degrees.
We have used these ELGs to determine the H$\alpha$ luminosity function (LF) of
actively star-forming galaxies and to derive the star-formation rate
density of the local universe.  We see substantial differences compared
to the previous work of Gallego \etal;
in particular, our value for L$^*$ is fainter than the Gallego \etal\
value, and our normalization is significantly higher.
By integrating our LF, we determine the total H$\alpha$
luminosity density for the  local universe and
the local star-formation  rate density. These values are
somewhat higher than those reported by Gallego \etal, indicating that the drop
off in star-formation activity between z of 1 and today was not as
severe as previously reported. 

\end{moriondabstract}

\section{Introduction}

The star formation rate (SFR) 
density of the local universe provides an important constraint
on models of galaxy formation and evolution.  In particular, 
measurements of the change in
global SFR history with redshift is critically
dependent on our knowledge of the local  SFR.  To date, this important 
number has only been measured by Gallego \etal\ \cite{Gallego}, with
a fairly small sample ($\sim 250$) of galaxies at low redshifts
($z_{lim} = 0.045$).  An independent check
of this value is needed.  

The KPNO International Spectroscopic Survey (KISS)
is a wide-field survey for extragalctic emission-line objects being carried
out with the Burrell Schmidt telescope at Kitt Peak.  A full description
of the survey can be found in \cite{Salzer}.  The main difference
between our survey and classical objective-prism searches for emission-line
galaxies (ELGs), such as that of Gallego \etal\ \cite{Gallego}, is our
use of a CCD detector which allows us to probe to much fainter 
magnitudes ($B \sim 20$) and higher redshifts ($z_{lim} = 0.085$) 
than previous photographic surveys.  As a result, our H$\alpha$
selected survey has detected 1126 ELG candidates in $\sim 68$
square degrees, for a surface density of 16.6 per square degree.
In contrast, the surface density of H$\alpha$ galaxies in the
Gallego \etal\ survey is 0.56 per square degree.

\section{Method}

Because our redshifts and H$\alpha$ fluxes were measured directly from the
low-dispersion objective-prism spectra, our
measurements are somewhat uncertain.
Currently our H$\alpha$ fluxes are calibrated using galaxies in common
with the survey of Gallego \etal\ \cite{Gallego}, and
in the absence of higher-resolution spectra, we have corrected for
internal reddening using
an empirical relation between $M_B$ and the Balmer decrement from the
emission-line galaxy survey of Salzer \cite{Salzer89}.  We are now
in the process of obtaining higher-resolution follow-up spectroscopy
for a subset of our galaxies.  These data will be used to improve
our redshift and flux calibration, but the conclusions stated below
should not be strongly affected.

We have measured the H$\alpha$ luminosity function of our
sample using H$\alpha$ luminosities derived from the measured
line fluxes and redshifts, and the $1/V_{max}$ estimator.  Note that
our sample is {\it not\/} magnitude limited, but is instead limited
by the line+continuum flux.  We therefore determined the completeness of the
survey using the $V/V_{max}$ test as outlined by Salzer\cite{Salzer89}.
This procedure yields 808 galaxies in our ``correctably complete'' sample. 

Our H$\alpha$ luminosity
function is shown in Figure 1 along with the H$\alpha$ luminosity function
measured by Gallego \etal\ \cite{Gallego}.  Note that our sample is sensitive
to substantially lower luminosities than that of Gallego.  The Schechter fit
parameters of our LF are: $L^* = 10^{41.85 \pm 0.04}$, $\phi^* = 0.00254\pm 0.0004$, and $\alpha = -1.18
\pm 0.07$.  Our $\phi^*$ is significantly higher than Gallego \etal\ who found
$\phi^* = 0.00063$.
 
The star formation rate (SFR) density of the local universe 
provides a critical constraint on our understanding of galaxy formation
and evolution.  The KISS sample allows us to determine this quantity
out to a redshift of 0.085. If we integrate the Schechter
luminosity function, we obtain a total $H\alpha$
luminosity density of 
\begin{equation}
L_{tot}  = \int_0^{\infty} \phi(L) L dL = \phi^* L^* \Gamma(2 + \alpha)
= 2.0 \times 10^{39}\ {\rm ergs~s}^{-1}~{\rm Mpc}^{-3}
\end{equation}
where $\Gamma$ is the gamma function.  
Since this H$\alpha$ luminosity reflects the number of ionizing
photons from massive stars, we can use this number to compute the
local SFR density.  If we adopt the conversion factor of
Madau \etal\ \cite{MPD} and assume a Salpeter initial mass function,
then
\[
L(H\alpha) = 1.5 \times 10^{41} \frac{SFR}{M_{\odot} \mathrm{yr^{-1}}} \mathrm{\ e
rgs \  s^{-1}}
\]
which implies a local SFR density of 0.013 $M_{\odot}$ yr$^{-1}$ Mpc$^{-3}$.
This value is a factor of
1.8 larger than that found by Gallego \etal\ \cite{Gallego}.
Our value for the local SFR density
of the universe is plotted in Figure 2 along with other measurements over
a range in redshift (\cite{Gallego}, \cite{Lilly}, \cite{Tresse}, 
\cite{Treyer}).  Note that the points derived from rest-frame UV measurements
(\cite{Treyer} and \cite{Lilly}) have {\it not\/} been corrected for
reddening.  The differences between the values derived using H$\alpha$ flux
and those using UV flux at the same redshifts can be used to place
constraints on the amount of internal extinction
present in star-forming galaxies.

\section{Summary}

We have measured H$\alpha$ luminosity function of the KISS ELG sample
and used this to determine the star formation rate density of the
local universe.  We find:

\begin{itemize}        
        \item the Schechter function parameters of the H$\alpha$ LF are:
$L^* = 10^{41.85 \pm 0.04}$, $\phi^* = 0.00254\pm 0.0004$, and $\alpha = -1.18
\pm 0.07$
       \item  the H$\alpha$ LF has a fainter 
characteristic luminosity ($L^*$) and a higher normalization ($\phi^*$)
than that of Gallego \etal\ \cite{Gallego}
        
	\item the integrated H$\alpha$ luminosity density
        of the local universe is $2.0 \times 10^{39}$ ergs s$^{-1}$ Mpc$^{-3}$.
	With the 
         assumption of a Salpeter IMF, this number implies 
	a star formation rate
        density of the local universe of 
	0.013 $M_{\odot}$ yr$^{-1}$ Mpc$^{-3}$.
        This is 1.8 $\times$ higher than that measured by Gallego \etal\
	\cite{Gallego}
       
 \end{itemize}
 Our value for the local star formation rate density greatly reduces
the observed evolution of star formation rate with redshift.

\acknowledgements

It is a pleasure to acknowledge my collaborators in the KISS project:
John Salzer, Alexei Kniazev, Valentin Lipovetsky,
Yuri Izotov, Todd Boroson, J. Ward Moody and
Trinh Thuan.  In addition, I would like to thank the many students who
aided in processing the survey data:  Laura Brenneman, Erin Condy, Mike
Santos, Karen Kinemuchi, Julie Barker, Janice Lee, Kathy Rhode, and Kristin
Kearns.  Valuable software support was provided by Jose Herrero and
Lisa Frattare.

%
\vfill\eject
\begin{figure}
\begin{center}
\leavevmode
\epsfverbosetrue
\epsfxsize=0.7\textwidth
\epsfbox{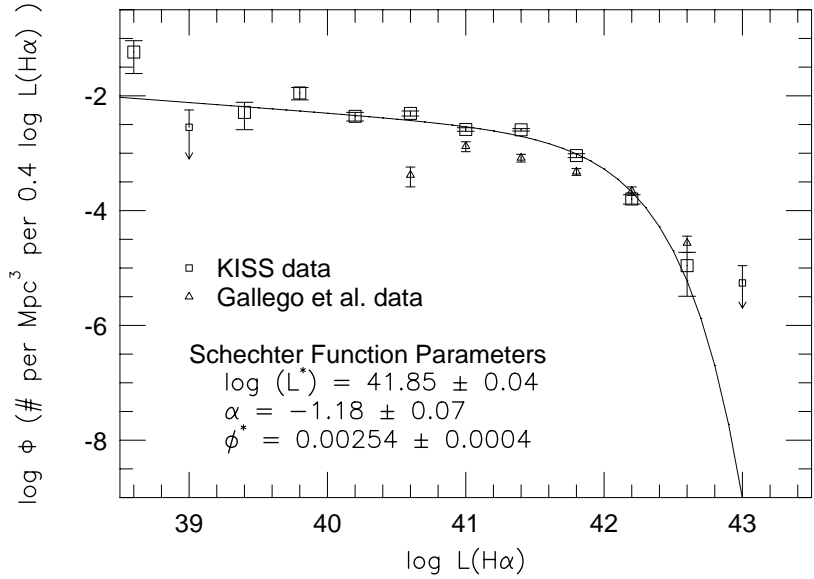}
\end{center}
\caption[]{$H\alpha$ luminosity function for the KISS emission-line
galaxies.  The solid line represents the Schechter function fit.
The error bars reflect the square root of the number of galaxies in each bin.  
Also shown is the $H\alpha$ luminosity function measured by \cite{Gallego}.}
\label{fig:lf}
\end{figure}

\begin{figure}
\begin{center}
\leavevmode
\epsfverbosetrue
\epsfxsize=0.7\textwidth
\epsfbox{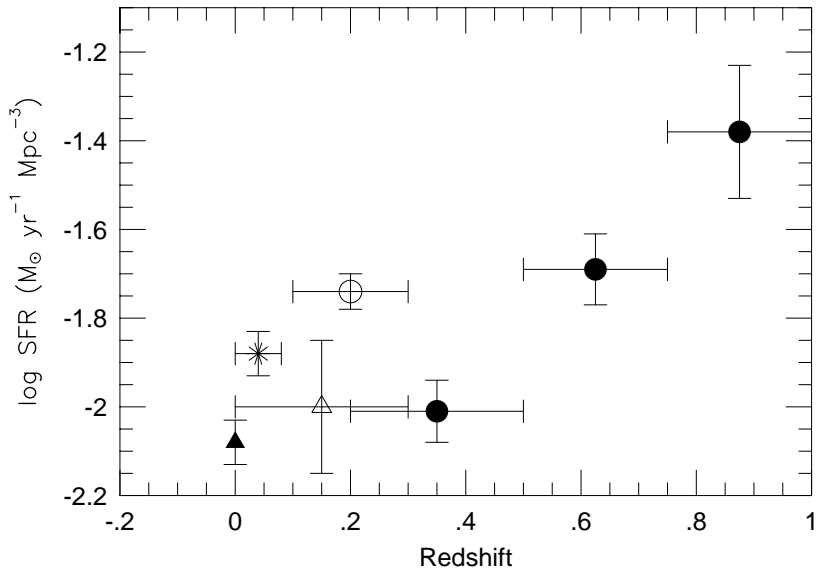}
\end{center}
\caption[]
{The observed redshift evolution of the star formation rate density
of the universe.  The asterisk represents our value, solid triangle is
from \cite{Gallego}, the open triangle is from \cite{Treyer},
the open circle is from \cite{Tresse}, and the solid circles
are from \cite{Lilly}.}
\label{fig:sfr}
\end{figure}
\end{document}